\def\vF{v_{\text{F}}}

\def\be{\begin{equation}}
\def\ee{\end{equation}}
\def\bea{\begin{eqnarray}}
\def\eea{\end{eqnarray}}
\def\bse{\begin{subequations}}
\def\ese{\end{subequations}}

\documentclass[prl,twocolumn,showpacs,preprintnumbers,amsmath,amssymb]{revtex4}
\usepackage{graphicx}
\usepackage{dcolumn}
\usepackage{bm}
\begin{document}
\preprint{}
\title{Quantum Correlations in Metals that Grow in Time and Space}
\author{T.R. Kirkpatrick$^{1}$ and D. Belitz$^{2}$}
\affiliation{$^{1}$Institute for Physical Science and Technology and Department
                   of Physics, University of Maryland, College Park, MD 20742\\
         $^{2}$Department of Physics and Theoretical Science Institute, University
                of Oregon, Eugene, OR 97403}
\date{\today}
\begin{abstract}
We show that the correlations of electrons with a fixed energy in metals have very anomalous time and space dependences. 
Due to soft modes that exist in any Fermi liquid, combined with the incomplete screening of the Coulomb interaction at finite 
frequencies, the correlations in 2-d systems grow with time as $t^2$. In the presence
of disorder, the spatial correlations grow as the distance squared. Similar, but in general weaker, effects are present in
3-d systems and in the absence of quenched disorder. We propose ways to experimentally measure these anomalous correlations.
\end{abstract}
\maketitle
Equilibrium time-correlation functions are an essential concept in statistical mechanics \cite{Forster_1975}. They describe the
spontaneous fluctuations of a system in equilibrium, and together with the partition function they provide a 
complete description of the equilibrium state. Via the fluctuation-dissipation theorem they also describe the
linear response of the system to external fields, and they are directly measurable by means of scattering
experiments. 

An old, and seemingly plausible, assumption is that microscopic correlations decay on time scales that are much
faster than macroscopic observation times. Various concepts 
depend on this assumption, for instance, the notion that the BBGKY hierarchy of classical kinetic equations
can be truncated \cite{Bogoliubov_1962}. An analogous assumption underlies the Kadanoff-Baym scheme of
deriving and solving quantum kinetic equations and its generalizations \cite{Kadanoff_Baym_1962, Langreth_Wilkins_1972}.
The assumption of a separation of time scales is also important in other areas, e.g., in signal
processing 
\cite{Schwinger_et_al_1998, Allen_Mills_2004}. For time-correlation 
functions it implies that they decay exponentially for large times. Equivalently, their Laplace transform is an analytic 
function of the complex frequency $z$ at $z=0$. The discovery of the non-exponential decay known as long-time tails (LTTs)
\cite{Alder_Wainwright_1967, Dorfman_Cohen_1970, Ernst_Hauge_van_Leeuwen_1970}, and the related
breakdown of a virial expansion for transport coefficients \cite{Dorfman_Cohen_1965, Weinstock_1965} thus 
came as a considerable surprise \cite{Peierls_1979}, since it showed that the assumption is in general not true. 
Rather, many time-correlation functions decay only algebraically, i.e., they have no intrinsic time scale. This scale 
invariance is reminiscent of the behavior of correlation functions at critical points; however, it occurs in entire phases, 
as opposed to just at isolated points in the phase diagram, and therefore is referred to as `generic scale invariance' 
\cite{Nagel_1992, Dorfman_Kirkpatrick_Sengers_1994, Belitz_Kirkpatrick_Vojta_2005}. The underlying physical 
reason is either conservation laws, or Goldstone modes that lead to a slow decay of
some long-wavelength fluctuations and, via mode-mode-coupling effects, affect the decay of other degrees of freedom. 
An example is the shear stress in a classical fluid, which is not conserved,
yet its time-correlation function decays algebraically as $1/t^{d/2}$ for long times $t$ in a $d$-dimensional fluid since it couples
to the transverse momentum, which is conserved. As a result, the Green-Kubo expressions for various
transport coefficients diverge in dimensions $d\leq 2$, and the hydrodynamic equations become nonlocal
in time and space; for a review, see Ref.~\onlinecite{Belitz_Kirkpatrick_Vojta_2005}.

In classical systems in equilibrium, LTT effects, while qualitatively very important, are rather small quantitatively
and become pronounced only at times so large that the correlation function is already very small overall. In
non-equilibrium classical systems the effects are much more important \cite{Kirkpatrick_Cohen_Dorfman_1982b, Ortiz_Sengers_2007}.
In equilibrium quantum systems the corresponding effects can also be much larger, especially in systems with quenched
disorder, where the quantum LTTs are often referred to as ``weak-localization effects'' \cite{Lee_Ramakrishnan_1985, 
Belitz_Kirkpatrick_Vojta_2005}. Still, the correlation functions considered to date decay as functions of time, albeit more slowly
than a separation-of-time-scales argument would suggest. In this Letter we show that in a quantum system as simple 
as interacting electrons with no quenched disorder, i.e., the simplest model of a metal, there are correlations that not 
only do not decay exponentially, but actually {\em grow} with time, and in some cases also with distance. This surprising result is a consequence of generic 
soft, or slowly decaying, excitations in a Fermi liquid in conjunction with the incomplete screening of the Coulomb 
interaction at nonzero frequencies. It is a dramatic illustration of the fact that the impossibility of separating
microscopic and macroscopic time scales, which is present in classical kinetics, holds {\em a fortiori} in quantum
systems. 

In quantum statistical mechanics it is useful to consider correlation functions that depend on one or more
imaginary-time variables $\tau \in [0,1/T]$, with $T$ being the temperature,
or on the corresponding imaginary Matsubara frequencies, $i\omega_n = 2i\pi T(n+1/2)$ for fermions, and
$i\Omega_n = 2i\pi Tn$ for bosons ($n$ integer). Functions defined for imaginary Matsubara frequencies can be
analytically continued to all complex frequencies, and the underlying real-time dependence can be obtained by
an inverse Laplace transform. The observables in a fermion systems can be expressed in terms of expectation
values of products of field operators ${\hat\psi}^{\dagger}({\bm x},\tau)$ and ${\hat\psi}({\bm x},\tau)$ that depend
on the position ${\bm x}$ in addition to $\tau$. Spin is not essential for our purposes, and we suppress 
it for now. Let us consider binary products of ${\hat\psi}^{\dagger}$ and ${\hat\psi}$, and an imaginary-time Wigner operator
${\hat W}({\bm X},{\bm x};{\cal T},\tau) =  {\hat\psi}^{\dagger}({\bm X}+{\bm x}/2,{\cal T}+\tau/2)\,{\hat\psi}({\bm X}-{\bm x},{\cal T}-\tau/2)$. 
In a field-theoretic formulation,
${\hat\psi}^{\dagger}({\bm x},\tau)$ and ${\hat\psi}({\bm x},\tau)$ correspond one-to-one to fermionic 
(i.e., Grassmann-valued)  fields ${\bar\psi}({\bm x},\tau)$ and $\psi({\bm x},\tau)$ \cite{Negele_Orland_1988}, 
in terms of which we define a Wigner field 
\bea
W({\bm X},{\bm x};{\cal T}\!,\tau) &=& {\bar\psi}({\bm X}+{\bm x}/2,{\cal T}+\tau/2)
\nonumber\\
&&\hskip 0pt \times\psi({\bm X}-{\bm x}/2,{\cal T}-\tau/2)
\label{eq:1}
\eea
in analogy to the operator
${\hat W}$. In common applications of real-time Wigner operators or fields, ${\bm X}$ and $\cal T$ correspond to the
``average'' or ``macroscopic" (presumed to be slow) length and time scale, and ${\bm x}$ and $\tau$ to the 
``relative'' or ``microscopic" (assumed to be fast) scales. The definition of the Wigner field reflects 
the assumption that it is possible and useful to 
separate these two scales \cite{Schwinger_et_al_1998, Kadanoff_Baym_1962}. In terms of it, the fluctuating particle number density is given by
$n({\bm X},{\cal T}) = W({\bm X},{\cal T};{\bm x=0},\tau=0)$, and the equilibrium single-particle Green function by
$G({\bm x},\tau) = \langle W({\bm X},{\cal T};{\bm x},\tau)\rangle$, where $\langle \ldots \rangle$ denotes an average
taken with the action governing the fermion system. If the average is taken in a non-equilibrium state, $\langle W\rangle$ also
depends on ${\bm X}$ and ${\cal T}$. In a real-time formalism, with macroscopic time $T$ and microscopic time $t$,
$\langle W({\bm X},{\bm x};T,t)\rangle = -i\,G^<({\bm X},{\bm x};T,t)$ is the
Green function $G^<$ defined in Ref.~\cite{Kadanoff_Baym_1962}. Its Fourier transform with respect to the microscopic
variables, $g^<({\bm X},T;{\bm p},\omega)$, is often interpreted as the density of particles with momentum ${\bm p}$ and energy
$\omega$ at the space-time point $({\bm X},T)$ \cite{Kadanoff_Baym_1962, Mahan_2000}. Switching back to imaginary
time and frequency, this identifies
\bea
\rho({\bm X},i\omega_n) &=& T\int_0^{1/T} d{\cal T}\,d\tau\,e^{i\omega_n\tau}\,W({\bm X},{\cal T},{\bm x}=0,\tau)
\nonumber\\
&=& \sum_{\sigma} {\bar\psi}_{n,\sigma}({\bm X})\,\psi_{n,\sigma}({\bm X})
\label{eq:2}
\eea
as the density of particles with energy $\omega_n$ at point ${\bm X}$, i.e., a spatial energy distribution. 
$\rho$ is related to the full 
phase-space distribution by summing over all momenta and averaging over the macroscopic (imaginary) time. Restoring 
spin, its spatial Fourier transform reads in terms of fermion fields
\bea
\rho({\bm k},i\omega_n) &=& \int d{\bm X}\ e^{-i{\bm k}\cdot{\bm X}}\,\rho({\bm X},i\omega_n)
\nonumber\\
&=& \sum_{{\bm p},\sigma} {\bar\psi}_{n,\sigma}({\bm p}+{\bm k}/2)\,\psi_{n,\sigma}({\bm p}-{\bm k}/2)\ ,\qquad
\label{eq:3}
\eea
where ${\bar\psi}_n({\bm p}) = \sqrt{T/V}\int dx\,e^{-ipx}\,{\bar\psi}({\bm x},\tau)$ and
$\psi_n({\bm p}) = \sqrt{T/V}\int dx\,e^{ipx}\,\psi({\bm x},\tau)$, with $px = {\bm p}\cdot{\bm x} - \omega_n\tau$,
$\int dx = \int d{\bm x}\int_0^{1/T}d\tau$, and $V$ the system volume. $\rho$ depends on a macroscopic 
wave vector ${\bm k}$, but a microscopic frequency $\omega_n$. This is in contrast
to the number density $n$, which depends on two macroscopic variables. To verify the physical interpretation of $\rho$
we note that its expectation value determines the density of states $N(\omega)$ via
\bse
\label{eqs:4}
\be
N(\omega) = \frac{-1}{\pi}\,\frac{1}{V}\,\text{Im}\langle\rho({\bm k}=0,i\omega_n \to \omega + i0)\rangle
\label{eq:4a}
\ee
The zeroth frequency moment gives the particle number $N$. With $\eta = 0^+$ the usual convergence factor
\cite{Fetter_Walecka_1971} and $n_{\text{F}}(\omega)$ the fermion distribution function we have
\be
N = T\sum_n e^{i\omega_n\eta} \langle\rho({\bm k}=0,i\omega_n)\rangle = V \int d\omega\,n_{\text{F}}(\omega)\,N(\omega)\ ,
\label{eq:4b}
\ee
and the first frequency moment gives the energy $E$ carried by the particles~\cite{energy_footnote},
\bea
E &=& T\sum_n e^{i\omega_n\eta} i\omega_n \langle\rho({\bm k}=0,i\omega_n)\rangle 
\nonumber\\
   &=& V \int d\omega\,n_{\text{F}}(\omega)\,\omega\,N(\omega)\ .
\label{eq:4c}
\eea
\ese

Let us now consider the four-fermion correlation function $C_{\rho\rho}({\bm X}-{\bm Y};i\omega_n,i\omega_m) = 
\langle \delta\rho({\bm X},i\omega_n)\, \delta\rho({\bm Y},i\omega_m)\rangle$, with $\delta\rho = \rho - \langle\rho\rangle$. 
This is motivated by two considerations. First, in light of the above interpretation of $\rho$, $C_{\rho\rho}$ provides
information about the correlations of energy levels in the Fermi system: It is the second moment of the energy density
distribution. Second, the quantity $\nu({\bm k},i\omega_n) = \rho({\bm k},i\omega_n) - \rho({\bm k},-i\omega_n)$
can be interpreted, in a technically precise sense, as an order parameter (OP) for the Fermi liquid~\cite{Belitz_Kirkpatrick_2012, OP_footnote}.
$C_{\nu\nu}({\bm k};i\omega_n,i\omega_m) = \langle\delta\nu({\bm k},i\omega_n)\,\delta\nu(-{\bm k},i\omega_m)\rangle $ is thus the 
(longitudinal) OP susceptibility in an ordered phase. We will come back to this interpretation below. Writing $C_{\rho\rho}$ 
in imaginary-time space, and using time translational invariance, one easily sees that it consists of two distinct contributions. One piece 
(which one would call ``disconnected'' in a diagrammatic representation) is proportional to $\delta_{nm}$, and a second, ``connected'' one, 
is proportional to $T$ \cite{T_factor_footnote}. We focus on the connected piece by putting $\omega_m = -\omega_n$ and eliminate the 
trivial factor of temperature by defining
\be
C({\bm k}, i\omega_n) = \frac{1}{VT}\,\langle\delta\rho({\bm k},i\omega_n)\,\delta\rho(-{\bm k},-i\omega_n)\rangle\ ,
\label{eq:5}
\ee
which has a well-defined zero-temperature limit \cite{EPL_footnote}. Since $C$ depends on a microscopic time scale, the separation-of-time-scales
assumption would suggest that the analytic continuation $C({\bm k},i\omega_n\to z)$ is an analytic function of the complex frequency 
$z$ at $z=0$, corresponding to exponential decay in imaginary or real frequency space. From ordinary LTT physics one might
expect that a coupling between the fast and slow degrees of freedom will lead instead to a nonanalytic function of the form $z^{\alpha}$,
which would lead to a LTT of the form $1/t^{\alpha + 1}$. We find that neither of these expectations is correct in general:  
In a Fermi liquid with a Coulomb interaction in $d=2$ the real-time dependence of the correlation
function $C$ is
\bse
\label{eqs:6}
\be
C({\bm k}\to 0, t) \propto \kappa^2 \ln(\kappa/\vert{\bm k}\vert)\, t^2 \qquad (d=2)
\label{eq:6a}                                                            
\ee   
where $\kappa$ is the screening wave number. That is, $C$ {\em increases} with time as $t^2$;
i.e., the correlations get {\em stronger} with increasing time. This behavior is cut off by a nonzero wave number $k$ or,
equivalently, by a finite system size $L$; it is valid for times $t \ll \sqrt{L\kappa}/\vF\kappa$ (and $t$ much larger than
the microscopic time scale), with $\vF$ the Fermi velocity. In $d=3$ the behavior is a LTT, 
\be
C({\bm k}\to 0, t) \propto 1/\vF^3\,t \qquad (d=3)\ ,
\label{eq:6b}                                                            
\ee   
\ese
which is valid for $t \ll L/\vF$. For asymptotically large times $C$ decays exponentially, with the rate of decay going
to zero as the wave number approaches zero or the system size goes to infinity.

Also of interest are the spatial correlations. For $i\omega_n \to 0$, i.e., for particles close to the Fermi surface, the
spatial correlations decay only algebraically,
\be
C({\bm x},i\omega_n \to 0) \propto \begin{cases} \sqrt{\kappa}/\vF^3\,\sqrt{r} & \qquad (d=2) \\
                                                                              \kappa/\vF^3\,r^2                & \qquad (d=3)
                                                                              \end{cases}
\label{eq:7}
\ee
for distances $r = \vert{\bm x}\vert \gg 1/\kappa$.  

These results hold for clean systems. In the presence of quenched disorder, the effects are even stronger.
The time dependence in $d=2$ is the same as in the clean case and given by Eq.~(\ref{eq:6a}), but
in $d=3$ the correlation function does not decay with time for $t \ll L^2/D$,
\be
C({\bm k}\to 0, t) \propto     1/D^2\,\vert{\bm k}\vert \qquad (d=3)\ ,
\label{eq:8}
\ee                                                          
where $D$ is the diffusion coefficient that characterizes the diffusive electron dynamics. The spatial
correlations for particles near the Fermi surface grow quadratically with the distance and remain
constant in $d=2$ and $d=3$, respectively,
\be
C({\bm x},i\omega_n \to 0) \propto \begin{cases} (1/D^3)\,r^2 & \qquad (d=2) \\
                                                                              1/D^3                & \qquad (d=3)\ .
                                                                              \end{cases}
\label{eq:9}
\ee
These expressions are again valid for distances large compared to the microscopic length. Note that for 
particles at the Fermi surface ($i\omega_n = 0$) the spatial correlations at large distances in $d=3$ 
get cut off only by the system size. In $d=2$ they grow as the square of the distance for distances
less than the localization length or the system size, whichever is smaller.

\begin{figure}[t]
\vskip -0mm
\includegraphics[width=8.5cm]{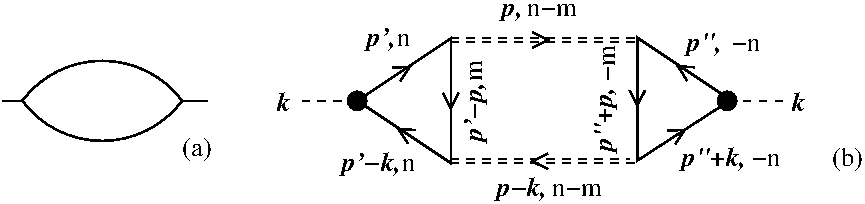}
\caption{Diagrammatic representation of the correlation function $C({\bm k},i\omega_n)$ for clean
              systems within
              (a) the effective field theory of Ref.~\cite{Belitz_Kirkpatrick_2012}, and (b) many-body
              perturbation theory. In (b), solid and double-dashed lines denote electronic Green 
              functions and dynamically screened Coulomb potentials, respectively. Notice that no
              frequency is transferred at the external vertices (heavy dots); this reflects the
              fact that this is not a contribution to the usual density correlation function. The
              frequency conservation at the internal vertices is as usual.}
               \vskip -0mm
\label{fig:1}
\end{figure}
We now sketch the origin and derivation of these surprising
results, and then discuss their physical significance as well as ways to check them experimentally.
We first consider clean systems.
The correlation function $C$, Eq.~(\ref{eq:5}), can be calculated in various ways. In the framework
of the effective field theory developed in Ref.~\onlinecite{Belitz_Kirkpatrick_2012} the leading
contribution is given by the one-loop diagram shown in Fig.~\ref{fig:1}(a). The advantage of this
framework is that the renormalization-group analysis of the effective theory guarantees that the
result is the leading behavior; higher-loop diagrams will change the prefactor, but not the 
functional form of the result. Alternatively, the same result can be obtained from
many-body perturbation theory \cite{Abrikosov_Gorkov_Dzyaloshinski_1963} via the diagram shown 
in Fig.~\ref{fig:1}(b); however, there is no such guarantee within that formalism. A simplified
analytical expression for either diagram, which has the correct scaling behavior, at $T\to 0$ is
\be
C({\bm k},i\omega_n) \propto \int_{k}^{\Lambda} dp\,p^{d-1} \int_{\omega_n}^{\infty} d\omega\,
   \frac{1}{(\omega^2 + \vF^2p^2)^2}\,\left(U(p,\omega)\right)^2.
\label{eq:10}
\ee
Here $\Lambda$ is an ultraviolet momentum cutoff, and $U(p,\omega)$ is the dynamically
screened Coulomb interaction. The factor of $1/(\omega^2 + \vF^2 p^2)^2$ represents the
soft fermionic modes. The strongly singular behavior discussed above results from
a combination of the latter and the incomplete screening of the Coulomb interactions at
nonzero frequencies. A short-range interaction still leads to singularities, but they are
weaker than in the Coulomb case; the corresponding behavior is obtained by
replacing $U(p,\omega)$ in Eq.~(\ref{eq:10}) by a constant. The limit on the time regime
where Eq.~(\ref{eq:6a}) is valid results from the most singular behavior of $C$ in $d=2$ being
restricted to frequencies $\omega_n$ larger than the plasma frequency. Note that the latter
can be made arbitrarily small by going to small wave numbers (or to large system sizes at
${\bm k}=0$). 

For disordered systems, an appropriate effective field theory is the generalized nonlinear
sigma model that has been studied extensively in the context of metal-insulator transitions
\cite{Finkelstein_1983, Belitz_Kirkpatrick_1997, Belitz_Kirkpatrick_1994}. The relevant one-loop diagram is still given
by Fig.~\ref{fig:1}(a), but the nature of the propagators is diffusive rather than ballistic. 
Within the framework of many-body perturbation theory the diagram shown in
Fig.~\ref{fig:1}(b) needs to be dressed with diffusion poles in elaborate ways. With any
calculation method the net result is that the factor $1/(\omega^2 + \vF^2 p^2)^2$ in
Eq.~(\ref{eq:10}) gets replaced by a diffusion pole to the fourth power, and the dynamically
screened Coulomb potential gets modified to reflect the diffusive nature of the electron
dynamics.

In summary, we have shown that a correlation function that describes particle-number
fluctuations with a fixed energy in a Fermi liquid is a very singular function of space and
time. In the zero-wave-number limit, and in a $2$-$d$ system, the correlations grow quadratically 
with time for times small compared to the inverse plasma frequency, and they decay only
as $1/t$ in $d=3$. For particles near the Fermi surface, the correlations decay only
algebraically with distance in clean systems, Eq.~(\ref{eq:7}). In the presence of quenched
disorder, they grow quadratically with distance in $d=2$ and remain constant in $d=3$,
Eq.~(\ref{eq:9}). Remarkably, this behavior is cut off only by the system size. In what follows
we add some discussion remarks to put this remarkable behavior in context.

(1) As mentioned after Eq.~(\ref{eq:4c}), the correlation function $C$ can be interpreted
as an OP susceptibility for the Fermi liquid. An interesting analogy in this
context is the corresponding OP susceptibility in a classical Heisenberg
ferromagnet. Due to a coupling between the longitudinal and transverse magnetization
fluctuations the longitudinal magnetic susceptibility $\chi_{\text{L}}$ (i.e., the OP susceptibility)
for $2<d<4$ diverges everywhere in the ordered phase as $1/k^{4-d}$ \cite{Brezin_Wallace_1973}. 
This results from a one-loop contribution to $\chi_{\text{L}}$ that is a wave-number convolution of 
two Goldstone modes, each of which scales as an inverse wave number squared. Diagrammatically
this contribution has the same form as Fig.~\ref{fig:1}(a). To see the origin of the stronger effects
discussed here, consider the spatial variation of $\chi_{\text{L}}$ as a function of the distance.
Setting all wave-number components except for $k_x$ equal to zero, ${\bm k} = (k_x,0,\ldots)$, we have
\be
\chi_{\text{L}}(\vert{\bm x}\vert \to \infty) = \int dk_x\,e^{ik_xx}\,\chi_{\text{L}}(k_x)
     \propto \vert x\vert^{3-d}\ .
\label{eq:11}
\ee
That is, the correlations grow with distance for $2<d<3$. Our results for the Fermi-liquid
OP susceptibility $C$ are in direct analogy to this result if one makes the
following adjustments: (i) For the time or frequency dependence, replace 
$k_x$ by the frequency and put ${\bm k}=0$.
(ii) Realize that the relevant propagator in the quantum field theory \cite{Belitz_Kirkpatrick_2012} scales as a soft
mode squared, see Eq.~(\ref{eq:10}). In the many-body calculation, this is apparent
from the triangular fermion loops in Fig.~\ref{fig:1}(b), each of which scales as a
ballistic propagator squared. 
(iii) Take into account the incomplete screening of the Coulomb interaction, which
enhances the effect compared to the naive expectation that the quantum result
should correspond to the classical one in an effective dimension $d_{\text{eff}} = d+1$.

(2) The temporal and spatial dependences of the correlation
function $C$ are quite different: The underlying correlation is a function of two points in space, but
four points in time; translational invariance implies that one and three of these, respectively,
are independent. The time dependence of the function $C$ results from having integrated
over two of the three independent time arguments, which is justified by the physical
interpretation of the function $C$. A related point is that we study the behavior of $C$ for
both frequency arguments approaching the Fermi surface, $\omega_n = -\omega_m \to 0$,
rather than for large frequency differences. The
spatial dependence of $C$, on the other hand, has the same structure as in
usual two-point correlation functions.  An important result is that the spatial correlations 
become more and more long ranged as the Fermi surface is approached. 

(3) In the classical-magnet analog the strong fluctuations eventually lead to an
instability of the ordered phase at the ferromagnetic transition. In the present case,
this suggests the possibility of a transition from a Fermi liquid to a non-Fermi liquid
with a vanishing density of states at the Fermi surface \cite{Kirkpatrick_Belitz_2012a}. However, whether such a
transition is actually realized in any given system is a question the current theory
cannot answer. 

(4) We suggest two ways to experimentally observe the effects discussed here: (i) A direct measurement
of the energy density distribution, e.g., in a cold-atom system, and (ii) measurements of the distribution of
the local density of states, e.g., by tunneling, which is related to the correlation function $C$ as
can be seen from Eq.~(\ref{eq:4a}).

(5) Studies of the distribution of the local density of states in disordered metals \cite{Lerner_1988, Andreev_Simons_Altshuler_1996}
have calculated a different correlation function:, viz., the disorder average of the
{\em disconnected} piece of the correlation function $C_{\nu\nu}$ defined after Eq.~(\ref{eq:4c}).
The effects considered were thus entirely determined by disorder fluctuations and vanish in the
clean limit. In contrast, the effects considered here are caused by the electron-electron interaction,
and some of them are further enhanced by disorder.


This work was supported by the NSF under grant Nos. DMR-1401410 and DMR-1401449. 
Part of this work was performed at the Aspen Center for Physics and supported by the
NSF under grants No. PHY-10-66293.


\begin{thebibliography}{35}
\bibitem[{T_f()}]{T_factor_footnote}
\bibinfo{note}{The disconnected piece can be written as the product of two
  Green functions, for the connected piece this is not possible. The factor of
  $T$ in the connected piece is a $T^2$ from the Fourier transforms of the
  fermion fields times a $1/T$ from a free imaginary-time integral. In a
  real-time formalism, the latter gets replaced by a free real-time integral
  which makes the correlation function proportional to an observation-time
  interval. $C$ defined in Eq.~(\ref{eq:5}) has a well-defined $T=0$ limit
  given by the corresponding real-time function normalized by the observation
  time.}
\end{thebibliography}

\end{document}